\documentclass[useAMS,usenatbib]{mn2e}
\usepackage{aas_macros}
\usepackage{graphics, graphicx}

\title[Variations of the amplitudes of oscillation of the Be star Achernar]{Variations of the amplitudes of oscillation of the Be star Achernar}
\author[K. J. F. Goss, C. Karoff, W. J. Chaplin, Y. Elsworth, I. R. Stevens]{K. J. F. Goss$^{1}$, C. Karoff$^{1,2}$, W. J. Chaplin$^{1}$, Y. Elsworth$^{1}$, I. R. Stevens$^{1}$\\
$^{1}$ School of Physics and Astronomy, University of Birmingham, Edgbaston, Birmingham, B15 2TT\\
$^{2}$ Department of Physics and Astronomy, Aarhus University, Ny Munkegade 120,DK-8000 Aarhus C, Denmark\\
Email: goss@bison.ph.bham.ac.uk}

\begin{document}

\pagerange{\pageref{firstpage}--\pageref{lastpage}} \pubyear{}

\maketitle

\label{firstpage}
\begin{abstract}
We report on finding variations in amplitude of the two main oscillation frequencies found in the Be star Achernar, over a period of 5 years.  They were uncovered by analysing photometric data of the star from the SMEI instrument.  The two frequencies observed, 0.775~d$^{-1}$ and 0.725~d$^{-1}$, were analysed in detail and their amplitudes were found to increase and decrease significantly over the 5-year period, with the amplitude of the 0.725~d$^{-1}$ frequency changing by up to a factor of eight.  The nature of this event has yet to be properly understood, but the possibility of it being due to the effects of a stellar outburst or a stellar cycle are discussed. 
\end{abstract}


\section{Introduction}

$\alpha$ Eridani, also known as Achernar (HD 10144), is one of the brightest stars in the Southern hemisphere.  With an apparent magnitude equal to 0.46, it is the brightest and one of the nearest Be stars to Earth \citep{2007NewAR..51..706K}.  

Be stars are non-supergiant B-type stars that show, or have shown at one time or another, emission in the Balmer line series.  The first Be star was reported in 1866 by Padre Angelo Secchi, where Balmer lines were observed in emission rather than in absorption \citep{2003PASP..115.1153P}.

For Be stars, the rotational velocity is 70-80$\%$ of the critical limit \citep{2003PASP..115.1153P}.  The rapid rotation causes two effects on the structure of the star: rotational flattening and equatorial darkening \citep{2007NewAR..51..706K}.  

Be stars have pulsation modes that are typical of $\beta$~Cephei and/or SPB stars, with frequencies roughly between 0.4~d$^{-1}$ (cycles per day) and 4~d$^{-1}$ \citep{2008CoAst.157...70G}.  A more complete review of Be stars may be found in \citet{2003PASP..115.1153P}.  

In this paper we present an analysis of the temporal variation of the two main oscillation frequencies detected in Achernar.  A description of the SMEI instrument used to collect the data is presented in Section 2.  An overview of the data analysis procedure is given in Section 3.  The results of the amplitude, frequency and phase analysis are presented in Section 4 and possible theories for the nature of the uncovered variation in oscillation amplitude are discussed in Section 5.  Finally, concluding remarks are in Section 6.

\section{SMEI}
Launched on 2003 January 6, the Solar Mass Ejection Imager (SMEI) on board the Coriolis satellite was designed primarily to detect and forecast Coronal Mass Ejections (CMEs) from the Sun moving towards the Earth.  However, as a result of the satellite being outside the Earth's atmosphere and having a wide angle of view it has been able to obtain photometric lightcurves for most of the bright stars in the sky.  These data have been used to study the oscillations of a number of stars, for example:   Arcturus \citep{2007MNRAS.382L..48T}, Shedir \citep{2009arXiv0905.4223G}, Polaris \citep{2008MNRAS.388.1239S}, $\beta$~Ursae Minoris \citep{2008A&A...483L..43T}, $\gamma$~Doradus \citep{2008A&A...492..167T}, $\beta$~Cephei stars (Stevens et al. 2010, in prep.) and Cepheid variables \citep{2010vsgh.conf..207B}.

SMEI consists of three cameras each with a field of view of 60$^{\circ}$ $\times$ 3$^{\circ}$, which are sensitive over the optical waveband. The optical system is unfiltered, so the pass band is determined by the spectral response of the CCD.  The quantum efficiency of the CCD is 45$\%$ at 700nm, falling to 10$\%$ at roughly 460nm and 990nm.  The cameras are mounted such that they scan most of the sky every 101 minutes, therefore the notional Nyquist frequency for the data is 7.086 d$^{-1}$.  Photometric results from Camera 1 and Camera 2 are used in the analysis of Achernar.  Camera 3 is in a higher temperature environment than the other two cameras and as a result the photometric data is highly degraded.  

\begin{figure}
\centering
\includegraphics[scale=0.65]{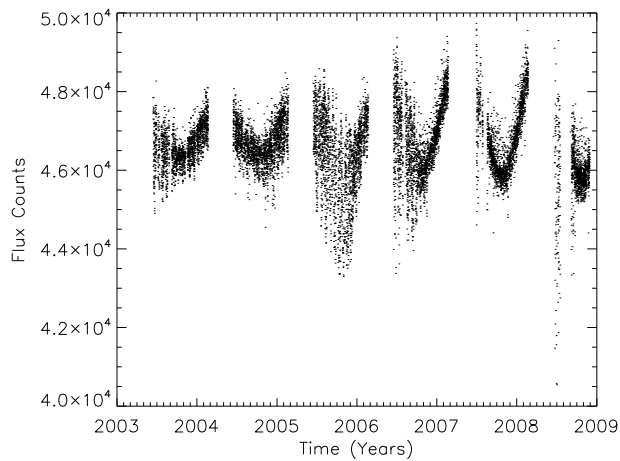}
\caption{5-year timeseries of Achernar data before a running mean was subtracted.}
\label{whole_timeseries}
\end{figure}

\begin{figure}
\centering
\includegraphics[scale=0.65]{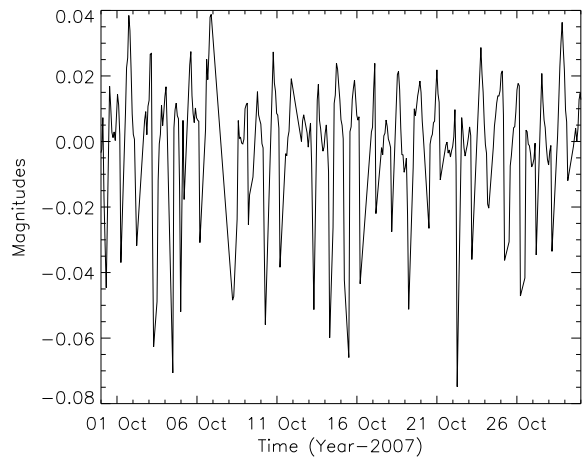}
\caption{30 day sample section of Achernar timeseries obtained with SMEI, which has been converted into magnitudes.}
\label{timeseries}
\end{figure}

The photometric timeseries for Achernar is shown in Figure~\ref{whole_timeseries}.  Note that the pronounced u-shapes in the lightcurve are due to effects from the SMEI instrumentation.  Since Camera 3 is not in use, the timeseries has a duty cycle of approximately 45$\%$.  This duty cycle is typical of most stars observed with SMEI, although for some stars the duty cycle can be considerably higher.  Figure~\ref{timeseries} shows an example segment of the Achernar timeseries obtained by SMEI where the flux has been converted into magnitudes.

SMEI is capable of detecting millimagnitude brightness changes in objects brighter than 6.5 magnitudes.  A detailed description of the SMEI instrument and the data analysis pipeline used can be found in (Spreckley $\&$ Stevens 2010, in prep.).

\section{Data Analysis}

A 5-year dataset for Achernar was obtained by SMEI, running from 2003 June 13 to 2008 November 26 (Figure~\ref{whole_timeseries}).  Long term variations in the data were removed by subtracting a running mean with a length of 10 days.  Various running mean lengths were tried and tested.  It was found that the choice of smoothing did not significantly affect the amplitudes or the frequencies being analysed, nor was the error on the smoothing significant enough to be included in the error analysis of the frequencies.  However the smoothing is required to reduce the noise at very low frequencies, e.g. long term variations in the timeseries such as the pronounced u-shapes in Figure~\ref{whole_timeseries}, an effect caused by the SMEI instrumentation.  The data were then converted into magnitudes for analysis (see Figure~\ref{timeseries} for example segment).

The timeseries as a whole was analysed using Period04 \citep{2005CoAst.146...53L}.  We used Period04 to analyse frequencies in the timeseries between 0.000~d$^{-1}$ and 7.086~d$^{-1}$, over which it uses a Discrete Fourier Transform algorithm to create an amplitude spectrum.

It is clear from the amplitude spectrum of Achernar (see Figure~\ref{fig:spec}), and other stars analysed using photometric data from SMEI, that there are frequencies present in the data that are due to the satellite. These frequencies occur at 1~d$^{-1}$, and multiples thereof, due to the sun-synchronous orbit of the satellite around the Earth.  Any genuine signals from the star around the 1~d$^{-1}$ frequencies cannot be distinguished from those caused by the orbit of the satellite and are disregarded in the analysis.

The timeseries, consisting of 1993 days in total, was then split into independent segments of 50 days.  Each individual segment was analysed using Period04 for the frequency and amplitude of the two main components detected in the spectra, at F1 (0.775~d$^{-1}$) and F2 (0.725~d$^{-1}$), where the aim was to search for temporal variations of the parameters.  Errors on the frequencies and amplitudes of the two main components were calculated using the Monte Carlo simulations in Period04 \citep{2005CoAst.146...53L}.  Changes in phase were calculated using Period04, whereby the phase in each 50 day period was calculated at a fixed frequency. 

To maintain consistency in the analysis between the different segments, only the F1 and F2 frequencies were pre-whitened in the amplitude spectrum. This meant that other significant frequencies may still have been in the timeseries, which may have had consequences for calculations such as the SNR (signal-to-noise ratio) (see Section~4).

\section{Results}

\subsection{New Frequencies Found}

\begin{figure}
\includegraphics[scale=0.65]{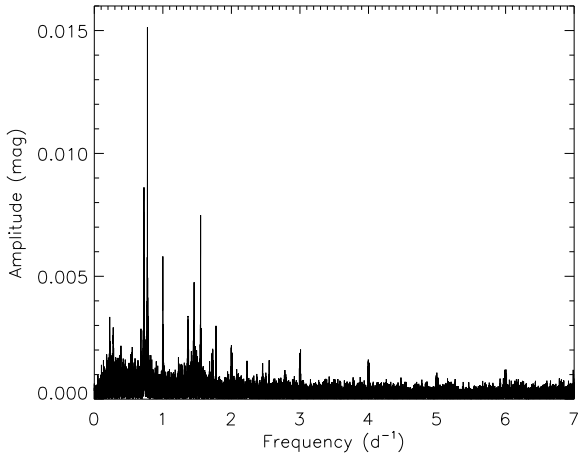}
\caption{Amplitude spectrum of Achernar, HD 10144}
\label{fig:spec}
\end{figure}

The amplitude spectrum of the 5-year dataset of Achernar can be seen in Figure~\ref{fig:spec}.  From this analysis, we are able to identify frequencies shown in Table~$\ref{tab:acher}$.  \citet{1987MNRAS.227..123B} first published a frequency of 0.792~d$^{-1}$ from simultaneous spectroscopy and photometry.  A slightly different frequency of 0.7745~d$^{-1}$ was then determined by \citet{2003A&A...411..229R} based on spectroscopic observations between 1996 and 2000 and is the more widely accepted value.  This is the frequency F1 (0.775~d$^{-1}$), in Table~\ref{tab:acher}. 

\citet{2006A&A...446..643V} reported on further frequencies using spectroscopic observations carried out between November 1991 and October 2000: 0.49~d$^{-1}$, 0.76~d$^{-1}$, 1.27~d$^{-1}$ and 1.72~d$^{-1}$.  Only evidence of the 0.76~d$^{-1}$ frequency, which is likely to be the same frequency reported in \citet{2003A&A...411..229R}, is evident in the SMEI data.  There does appear to be a group of frequencies around 1.72~d$^{-1}$ in the SMEI data (see Figure \ref{fig:spec}) but these were found to be combinations of the frequencies found in Table~\ref{tab:acher} and the 1~d$^{-1}$ frequency from the satellite.

Frequencies F2 (0.725~d$^{-1}$) and F3 (0.680~d$^{-1}$) are frequencies where no published results were found in the literature.  It is possible that the 1.72~d$^{-1}$ frequency observed by \citet{2006A&A...446..643V} is actually the frequency F2 observed with SMEI but with an additional 1 day cycle effect.  Further frequencies were found in the data, but these were the result of combinations of the frequencies mentioned above.  

\citet{2008CoAst.157...70G} reported on the first results on the Be stars observed with COROT.  They found that in one Be star non-sinusoidal signals were present after already removing approximately 50 frequencies suggesting that the amplitudes or frequencies of the signals were changing during the observations.  This is something that was observed when pre-whitening the data for Achernar in the initial amplitude spectrum.  Many frequencies around the F2 frequency were removed from the timeseries, but evidence of this signal still remained, hence providing evidence for linewidth.  This also occurred with the F1 frequency, but to a much lesser extent.

\begin{table}
\centering
\begin{tabular}{l l l l l}
\hline
& Frequency & Amplitude & SNR \\
& (d$^{-1}$) & (mag) & \\ 
\hline
F1 & 0.775177(5) & 0.0165(3) & 27.09\\
F2$^{\star}$ &0.724854(6) & 0.0129(3) & 19.05\\
F3$^{\star}$ & 0.68037(3) & 0.0027(3) & 4.11\\
\hline
\end{tabular}
\caption{Frequencies identified in Achernar, HD 10144.  The starred ($^{\star}$) frequencies represent frequencies where no published results were found in the literature.  Note: these frequencies are frequencies for the entire timeseries.}
\label{tab:acher}
\end{table}

\subsection{Amplitude variation}

\begin{figure}
\centering
\includegraphics[scale=0.65]{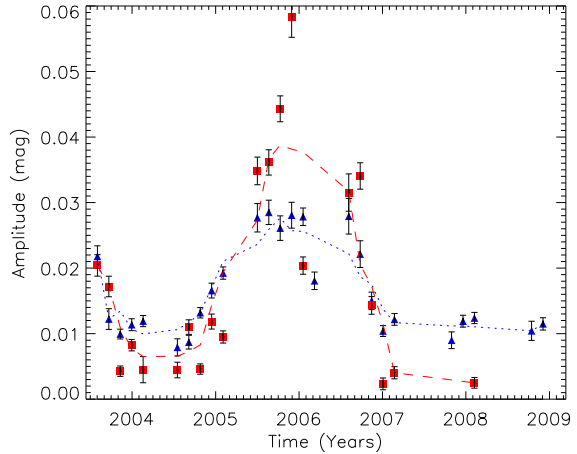}
\caption{A graph to compare the amplitude change of the two frequencies F1 and F2.  The blue triangles represent the F1 frequency and the red squares represent the F2 frequency.  The blue dotted line shows a smooth fit throught the F1 data points.  The red dashed line shows a smooth fit through the F2 data points.}
\label{fig:acher_amp}
\end{figure}

\begin{figure}
\includegraphics[scale=0.65]{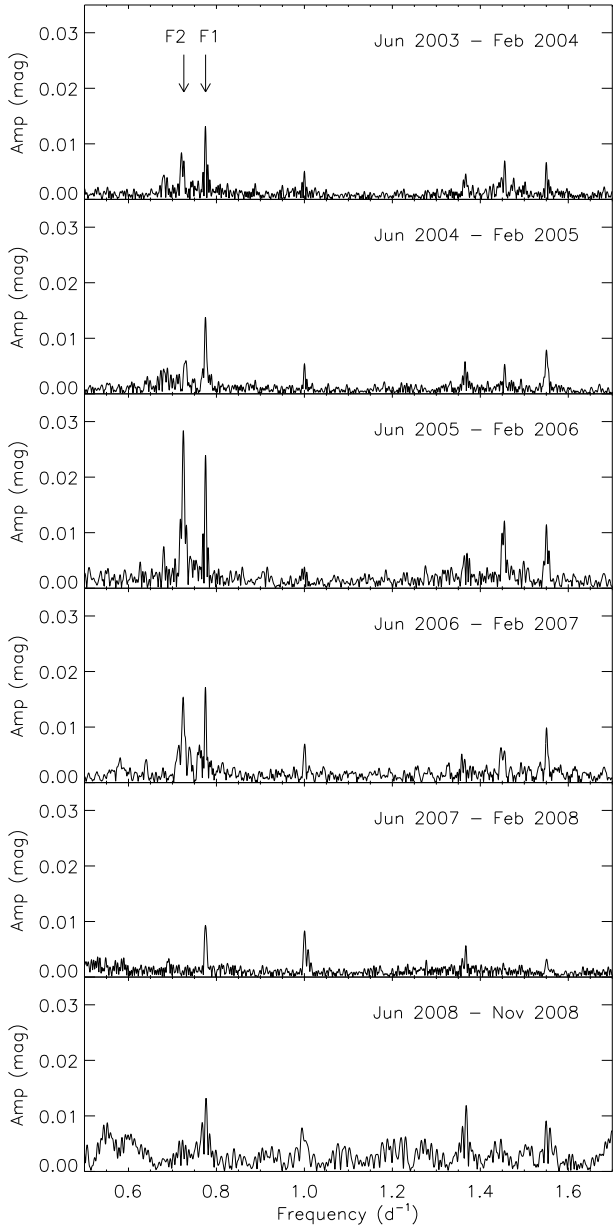}
\caption{Six amplitude spectra of Achernar at different epochs during the 5-year observation, showing frequencies between 0.5~d$^{-1}$ and 1.7~d$^{-1}$.  Note F2 disappears in bottom two panels.}
\label{fig:timepanel}
\end{figure}

The 5-year dataset was split into 50 day segments and the two frequencies with the largest amplitudes, F1 (0.775~d$^{-1}$) and F2 (0.725~d$^{-1}$), were analysed for changes in their frequency and/or amplitude.

Figure \ref{fig:acher_amp} shows a plot of the amplitudes of these two frequencies as a function of time.  The amplitudes vary and there is a significant increase in the amplitudes of both frequencies during the same time period, roughly between October 2004 and January 2007.  The F2 frequency starts with a lower amplitude than the F1 frequency, but during the period when the amplitudes increase, the amplitude of the F2 frequency increases above the amplitude of the F1 frequency.  The amplitudes of both frequencies decrease around January 2007, with the F2 frequency decreasing to an undetectable level, while the F1 frequency is still present.  The absence of the F2 frequency at this time is not through lack of points in the dataset.  The F2 frequency can no longer be detected in the 50 day time segments starting at: 2006-01-18, 2007-09-10, 2007-10-30, 2008-08-25 and 2008-10-14. 

The change in amplitude of the two frequencies is evident in Figure~\ref{fig:timepanel}, which shows the amplitude spectra of Achernar at six different epochs separated by the large gaps seen in Figure~\ref{whole_timeseries}.  Here it is obvious that the amplitudes of both frequencies increase, with the F2 frequency increasing significantly more than the F1 frequency, and then decreasing to an undetectable level at the end of the observation.  The noise around the frequencies increases when the amplitude increases. This can be seen when comparing the two top panels with the two middle planes in Figure 5. The fact that the noise around the frequencies increases when the amplitude increases suggests that the signals causing the frequencies may not be strictly coherent over timescales of hundreds of days. A non-coherent signal would also cause random phases. We therefore proceed to analyze frequency and phase variations in Section 4.3 below.

In order to rule out the increase in amplitude of the two frequencies being due to effects from the SMEI instrument we looked at variations in the oscillations and light curves of other stars for comparison.  In total, nine stars observed with SMEI were analysed to look for similar changes in the amplitude of oscillation, if oscillations were observed, and in the stability of the lightcurve over the same time period.  If the effect were dependent on Right Ascension and Declination then other stars in the vicinity of Achernar would show this trend.  Three stars in the vicinity of Achernar were analysed: HD 32249, HD 12311 and HD 3980, none of which showed the increase in amplitude.  Another possibility is that the increase in amplitude may only be obvious in very bright stars (Achernar being the 9th brightest star in the sky).  Arcturus, Vega and Capella (all stars brighter than Achernar) were analysed but no similar patterns were found.  Three stars of photometric reference were also analysed: HD~168151, HD~155410 and HD~136064 \citep{NeilTarrant:2010}, and they also showed null results.

\subsection{Frequency and Phase variation}

\begin{figure}
\centering
\includegraphics[scale=0.65]{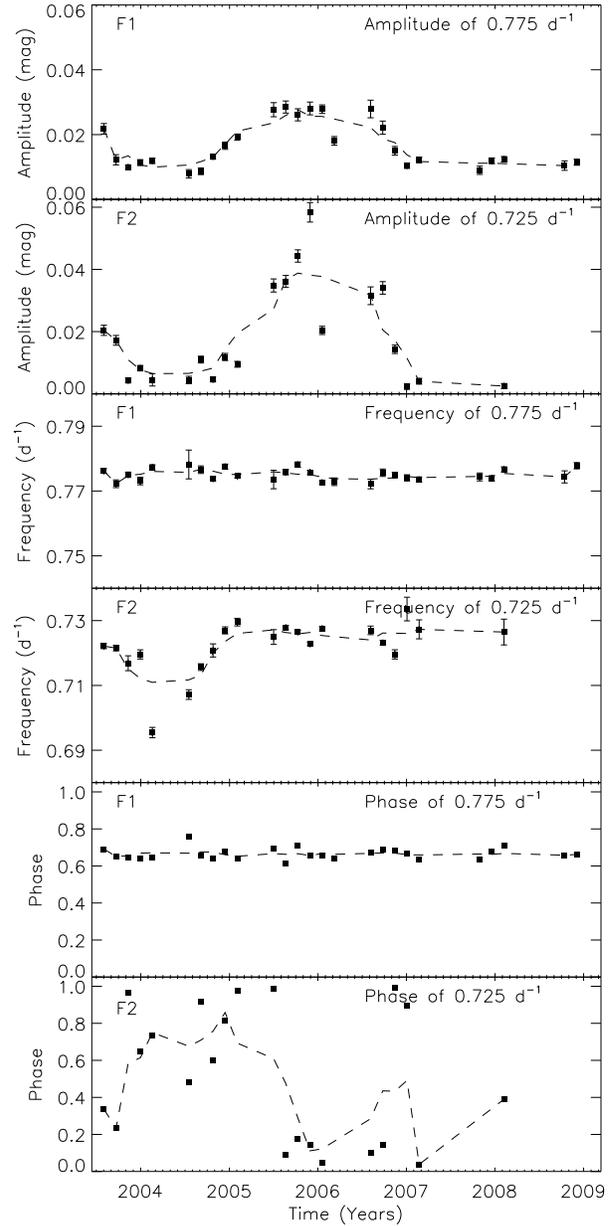}
\caption{$\emph{Panel 1:}$ Amplitude variations of the F1 frequency. $\emph{Panel 2:}$ Amplitude variations of the F2 frequency. $\emph{Panel 3:}$ Frequency variations in the F1 frequency. $\emph{Panel 4:}$ Frequency variations in the F2 frequency. $\emph{Panel 5:}$ Phase variations of the F1 frequency. $\emph{Panel 6:}$ Phase variations of the F2 frequency.  The dashed lines show a smooth fit through the data points.  Note that errors on some of the panels are smaller than the symbols.}
\label{fig:panel_amp_freq}
\end{figure}

Variations in Be stars can be ascribed to either rotation or non-radial oscillations.  It is generally assumed that the oscillations will have constant frequency and phase whereas rotationally modulated variations will have a transient nature and thus non-constant frequency and phase i.e. they will be non-coherent.  Both the frequency and phase of what are believed to be rotationally modulated variations can change due to outbursts from the central star to the surrounding disc (see \citet{2003A&A...411..167S} for a discussion of this).  On the other hand \citet{2009A&A...506...95H} saw similar amplitude changes in relation to an outburst in what they believed were non-radial oscillations.

The observations that we present here cover 5 years and thus are expected to cover many outbursts, but we do not have any information when these outbursts have taken place.  Also the time scales of the amplitude changes that we report here are much longer than the expected time scale of the outbursts.  We do therefore not have the possibility to correlate individual outbursts with amplitude, frequency or phase changes.  On the other hand if the frequencies and phases of the identified oscillations are indeed coherent over the 5 year time-span it would seem likely that the variability is due to oscillations.

In Figure~\ref{fig:panel_amp_freq} it is seen that F1 is a coherent oscillation with constant frequency and phase over the 5 year time-span. F2 shows a decrease in its frequency during 2004 and what appears to be a random phase i.e. it is not fully coherent. This makes it possible that F2 is due to rotational modulation, but the oscillation scenario cannot be completely ruled out.  Firstly, the similarity of the amplitudes of F1 and F2 suggest a common origin.  Secondly, it is not obvious that the lifetimes of the oscillations in Be stars are long compared to the 50 day segments used in this analysis.  And thirdly, the change in the frequency of F2 appears at low amplitude and thus a low S/N.  It is therefore not clear if the frequency change is indeed significant.

\section{Discussion}

An explanation of the nature of the observed amplitude variation could be a transient frequency during a stellar outburst as explained by \citet{2003A&A...411..229R} who report on non-radially pulsating Be stars.  Be stars are known for their stellar outbursts where a large transfer of mass from the star to its circumstellar disc occurs.  \citet{2003A&A...411..229R} discuss transient periods that are within 10$\%$ of the main photospheric period and which only appear during outburst events.

It is possible that the change in amplitude of the frequencies is due to temporary changes in the surface of the star such as a stellar outburst.  \citet{2009A&A...506...95H} report on the analysis of the Be star HD 49330, observed with the CoRoT satellite.  They find a direct correlation between amplitude variations in the pulsation modes and outburst events.  The amplitudes of the main frequencies (p mode oscillations, where gradients of pressure are the dominant restoring force) decrease before and for the duration of the outburst, only increasing after the outburst has finished.  Other groups of frequencies (g mode oscillations, where gravity is the dominant restoring force) appear just before the outburst reaching maximum amplitudes, during the outburst and then disappearing once the outburst is over.  However, it had not been determined whether the variations in pulsation modes produced the outburst, or whether the outburst leads to the excitation of the pulsation modes.  

\citet{2009A&A...506...95H} show that the changes in stellar oscillations from possible stellar outbursts last up to tens of days whereas the change in amplitude of the frequencies in Achernar last much longer, up to approximately 1000 days.  Long term variations in Be stars that last from months to years have been attributed to structural change in the circumstellar disk, e.g. an outburst filling the circumstellar disk with new material \citep{2009CoAst.158..194N}.  However, the longer duration may be an indication that the variations we observe are not linked to an outburst event, but relates more to the internal structure of the star and could be evidence for a cycle similar to the Sun's solar cycle.  

In the Sun the frequencies and amplitudes of the acoustic modes show variations that follow the changing magnetic activity during the solar cycle \citep{1990Natur.345..322E}.  For the low-degree modes, the fractional change in frequency is approximately 1.3 $\times$~10$^{-4}$ and the fractional change in amplitude approximately 0.2.  Given a cycle effect for Achernar that changes both amplitude \emph{and} frequency and also making the crude assumption that the ratio of the fractional changes in amplitude and frequency are the same as for the Sun, we find that we do not have the precision to detect such a change in frequency.  Even if the cycle were to only change the amplitude, resulting in the associated amplitude modulation mentioned in Section 4.3, the frequency change is still too small to be seen.

\citet{2006A&A...446..643V} found long term variations of the equivalent width of the H$\alpha$ line in Achernar.  These variations show that Achernar was in a strong emission phase (or Be phase) around 1965, 1978 and 1994.  If the oscillation amplitude changes presented here are related to a B to Be phase transition then the changes suggest that Achernar was in a Be phase around 2006.  Though we are not aware of any reports of Achernar showing strong emission around 2006, such a scenario is inconsistent with the 14-15 year cyclic B to Be phase transition suggested by \citet{2006A&A...446..643V}.

\citet{2008A&A...484L..13K} found that the orbital period of the close companion of Achernar was approximately 15 years and as a result its periodicity could be the trigger of the Be episodes.  Again, if correct, this would imply that Achernar would be in a Be phase around 2010 whereas the oscillation amplitude variations indicate 2006.

\section{Conclusions}

The long duration of the SMEI photometric data has allowed us to study the variations in the pulsation modes of the Be star Achernar over a period of 5-years.  

Analysis of the complete 5-year dataset has uncovered three significant frequencies: F1 (0.775~d$^{-1}$), F2 (0.725~d$^{-1}$), F3 (0.680~d$^{-1}$), of which only F1 has been published previously.  F2 is believed to be transient in nature from analysis of the independent time segments, a phenomenon that the SMEI instrument has the ability to detect due to its long photometric timeseries.  F3 has a SNR close to four and this frequency may be a pulsation or transient frequency.

Analysis of the independent time segments showed that the amplitudes of the two main frequencies, F1 and F2, have significantly increased and then decreased over the period of 5-years.  As discussed, this may be explained by the presence of a stellar outburst or a stellar cycle, but for the present these speculations remain inconclusive. 

\section*{Acknowledgments}
We would like to thank Steven Spreckley for his work in developing the SMEI pipeline and also Neil Tarrant for his useful input.  We also thank Coralie Neiner for fruitful discussions.

K.J.F.G., W.J.C., Y.E. and I.R.S. acknowledge the support of STFC.  C.K. acknowledges the support from the Danish Natural Science Research Council.

\bibliographystyle{mn2e}

\bibliography{redo_achernar}

\appendix

\label{lastpage}

\end{document}